\documentclass{aa}  
\usepackage{natbib}
\bibpunct{(}{)}{;}{a}{}{,} 
\usepackage{xspace}
\usepackage{mathrsfs}
\usepackage[free-standing-units = true,
space-before-unit = true,
use-xspace = true,
per-mode = power-positive-first]{siunitx}
\usepackage{multicol}
\usepackage[switch,columnwise]{lineno}
\usepackage{printlen}

\DeclareSIUnit \msun {\text{\ensuremath{M_\odot}}}

\DeclareSIUnit \kms {\kilo \meter \per\second}

\DeclareSIUnit \myr {Myr}

\DeclareSIUnit \kk {\kilo \kelvin}

 \def\mso{\,\mathrm{M}_\odot}
 \def\simgr{\mathrel{\hbox{\rlap{\hbox{\lower4pt\hbox{$\sim$}}}\hbox{$>$}}}}

\usepackage{graphicx}
\usepackage{txfonts}
\usepackage[colorlinks,citecolor=blue,linkcolor=blue]{hyperref}
\usepackage{hyperref}
\makeatletter
\renewcommand*\aa@pageof{, page \thepage{} of \pageref*{LastPage}}
\makeatother

\graphicspath{{figs_new/}}
\makeatletter
\newcommand\thefontsize[1]{{#1 The current font size is: \f@size pt\par}}
\makeatother
\usepackage{xcolor}
\definecolor{ochre}{rgb}{0.8, 0.47, 0.13}
\definecolor{barbiepink}{rgb}{.8784, 0.1294, 0.5412}

\begin{document}

	
	\title{Spindown of massive main sequence stars in the Milky Way}
	
	
	\author{
		K. Nathaniel\thanks{kjn9240@rit.edu}\inst{1,2}
		\and
		N. Langer\inst{1}
		\and
		S. Sim\'on-D\'iaz\inst{3,4}
		\and 
		G. Holgado\inst{3,4}
		\and
		A. de Burgos\inst{3,5}
		\and
		B. Hastings\inst{2}
	}
	
	\institute{
		Argelander-Institut f\"ur Astronomie, Universit\"at Bonn,
		Auf dem H\"ugel 71, 53121 Bonn, Germany
        \and
        Center for Computational Relativity and Gravitation, Rochester Institute of Technology, Rochester, NY 14623 USA
		\and
		Instituto de Astrof\'isica de Canarias, 38200 La Laguna, Tenerife, Spain
		\and
		Departamento de Astrof\'isica, Universidad de La Laguna, 38205 La Laguna, Tenerife, Spain
        \and
        European Southern Observatory, Alonso de C\'ordova 3107, Vitacura, Santiago, Chile\\
	}
	
	\date{Received XXX; accepted YYY}
	
	
	\abstract
	{We need to understand the spin evolution of massive stars to compute their internal rotationally induced mixing processes, isolate effects of close binary evolution, and predict the rotation rates of white dwarfs, neutron stars and black holes.}
	{We discuss the spindown of massive main sequence stars imposed by stellar winds.}
	{We use detailed grids of single star evolutionary models to predict the distribution of the surface rotational velocities of core-hydrogen burning Galactic massive stars as function of their mass and evolutionary state. We then compare the spin properties of our synthetic populations with appropriately selected sub-samples of Galactic main sequence OB-type stars extracted from the IACOB survey.
    }
	{We find that below \qty{\sim 40}{\msun}, observations and models agree in finding that the surface rotational velocities of Galactic massive stars remain relatively constant during their main sequence evolution. The more massive stars in the IACOB sample appear to spin down less than predicted, while our updated angular momentum loss prescription predicts an enhanced spindown. 
    Furthermore, the observations show a population of fast rotators, with $v \sin i \simgr \qty{200}{\kms}$ persisting for all ages, which is not reproduced by our synthetic single star populations. } 
	{
	We conclude that the wind-induced spindown of massive main sequence stars is yet to be fully understood, and that close binary evolution might significantly contribute to the fraction of rapid rotators in massive stars. }
	
	\maketitle
	%
	
\section{Introduction} \label{sec:intro}

Rotation plays a key role for the whole evolution of massive stars, from their birth to the formation of their end products. Indeed, models predict that it may have an impact as strong as their initial mass or metallicity \citep{meynetStellarEvolutionRotation2000,langerPresupernovaEvolutionMassive2012,deminkRotationRatesMassive2013}.
As a massive star evolves, its surface equatorial velocity is influenced by several factors.
Massive single stars typically spin down, due to the coupling of angular momentum loss by stellar winds and efficient internal transport of angular momentum \citep{langerCoupledMassAngular1998}.
This is not true for stars in binary systems, where interactions such as mass transfer can cause a star to increase or decrease its rotation rate \citep[e.g.,][]{zahnDynamicalTideClose1975,packetSpinupMassAccreting1981,petrovicWhichMassiveStars2005,deminkRotationRatesMassive2013}.

Understanding stellar rotation requires both deriving projected rotational velocities ($v\sin i$) from spectroscopic data \citep[e.g.,][]{contiSpectroscopicStudiesOtype1977, pennyProjectedRotationalVelocities1996, huangStellarRotationCensus2010, zorecRotationalVelocitiesAtype2012, ramirez-agudeloVLTFLAMESTarantulaSurvey2013, markovaSpectroscopicPhysicalParameters2014, simon-diazIACOBProjectRotational2014, duftonCensusMassiveStars2019, daherStellarMultiplicityStellar2022,sunExploringStellarRotation2024} as well as studying how rotation impacts stellar structure and evolution. The former can be used to constrain the initial rotation rate distribution \citep{rosenWhatSetsInitial2012, holgadoIACOBProjectVII2022}, investigate the spins of evolved magnetic stars \citep{fossatiEvidenceMagneticField2016, petitMiMeSSurveyMagnetism2019,keszthelyiEffectsSurfaceFossil2020}, study rotationally induced mixing and chemically homogeneous evolution \citep{yoonEvolutionRapidlyRotating2005, martinsEvidenceQuasichemicallyHomogeneous2013, martinsStudyEffectRotational2017,markovaSpectroscopicPhysicalParameters2018}, probe angular momentum loss \citep{granadaEvolutionSingleBtype2014, rieutordDynamicsRadiativeEnvelope2014,gagnierCriticalAngularVelocity2019}, diagnose magnetic braking \citep{meynetMassiveStarModels2011, curtisTemporaryEpochStalled2019, hallWeakenedMagneticBraking2021, davidFurtherEvidenceModified2022}, assess theories of massive star formation \citep{wolffStellarRotationClue2006, mokiemVLTFLAMESSurveyMassive2006, huangStellarRotationCensus2010,chiappiniImprintsFastrotatingMassive2011}, and constrain rotation rates of final compact objects \citep{hegerPresupernovaEvolutionRotating2000, petrovicWhichMassiveStars2005}.
Most work on stellar wind mass loss, including this study, has been done with one-dimensional (1D) models, but rapidly rotating stars are distorted from spherical symmetry due to both centrifugal forces and the fact that flows across and in the star are two-dimensional \citep{friendTheoryRadiativelyDriven1986, pelupessyRadiationDrivenWinds2000, mullerRotatingMassiveStars2014, hastingsModelAnisotropicWinds2023}, which in the end also produces effects in the stellar observables \citep{abdul-masihEffectsRotationSpectroscopic2023}.

Stars are generally thought to be born spinning rapidly, due to the large angular momentum content of their birth clouds. While solar type stars spin down dramatically due to the coupling of their winds with their convectively generated magnetic fields \citep{skumanichTimeScalesCa1972,davidFurtherEvidenceModified2022}, massive main sequence (MS) stars, on the other hand, are thought to spin down much less, even though their winds are stronger by many orders of magnitude. Besides their short life time, this is because large scale surface magnetic fields are mostly absent, with the few percent which show magnetic fields also undergoing a strong spindown \citep{fossatiEvidenceMagneticField2016,petitMiMeSSurveyMagnetism2019}.

In non-magnetic massive single stars, three effects influence the evolution of their surface rotation velocity during the MS: (i) expansion during the MS, (ii) internal angular momentum redistribution, and (iii) angular momentum loss due to winds.

Massive stars on the MS expand by a factor of \num{\approx 3}, for which angular momentum conservation would imply a corresponding spindown of the stellar surface. However, the envelope expansion is accompanied by simultaneous core contraction and efficient transport of the resulting excess angular momentum from the core to the envelope \citep{ekstromEvolutionCriticalLimit2008, brottRotatingMassiveMainsequence2011a, deminkRotationRatesMassive2013}.
These two processes have opposite effects on the surface rotation, with the latter being more relevant for massive stars, due to their larger core mass fraction \citep{hastingsSingleStarPath2020}.

Non-magnetic stars also lose angular momentum in the case that they undergo stellar wind mass loss.
In a simplified picture of spindown due to stellar winds, the mass removed by stellar winds also contains angular momentum. The new surface layers expand and would spin down if their angular momentum was conserved. However, as any shear would induce angular momentum transport towards reestablishing rigid rotation, angular momentum from deep inside is brought towards the surface to compensate. The net effect is a spindown of the whole star. The dominant effect of a stellar wind on a non-magnetic rotating star is indeed angular momentum loss rather than mass loss, as the relative rate of angular momentum loss exceeds that mass loss by one order of magnitude \citep{langerCoupledMassAngular1998}.

On the MS, the winds are stronger at higher masses and higher metallicities \citep[e.g.,][]{kudritzkiWindsHotStars2000,mokiemEmpiricalMetallicityDependence2007}. For massive Milky Way stars, angular momentum loss is expected to have an important influence on spindown for MS stars above \qty{\sim 9}{\msun} \citep{langerPresupernovaEvolutionMassive2012}.

In this study, we extend the work of \citet{holgadoIACOBProjectVII2022} to
investigated the spindown of Galactic massive stars across the MS.
We derive predictions for the distribution of the surface rotational velocities of MS Galactic massive stars based on this simple picture, and compare them to the observed evolution of the rotation rate as obtained from the IACOB survey \citep{simon-diazIACOBProjectSynergies2011, simon-diazIACOBSpectroscopicDatabase2020}.
In Sect.\,\ref{sec:models} we discuss our evolutionary models and describe the population synthesis algorithm developed for this paper. In Sect.\,\ref{sec:model_predictions} we analyze the spindown evolution of our models. In Sect.\,\ref{sec:obs_sample} we discuss the sample of observed stars. In Sect.\,\ref{sec:compare_to_obs} we compare the distributions of projected rotational velocity of the observed stars and the synthetic populations with two different angular momentum prescriptions. In Sect.\,\ref{sec:discuss} we discuss the uncertainties in our methods and the observed sample. We provide a brief summary in Sect.\,\ref{sec:conclusions}.

\section{Method} \label{sec:models}
We use the solar metallicity evolutionary models of \cite{brottRotatingMassiveMainsequence2011a} to derive predictions for the evolution of the surface rotational velocity of massive stars across the MS. We first describe the essential physics assumptions in these models, then discuss the assumptions on stellar wind induced angular momentum loss. We do this for the original models, and we describe how, based on the data from the original models, we use a newly devised angular momentum loss prescription. Furthermore, we explain the way we use the model data to derive a synthetic population of massive stars in order to predict the expected distribution of rotational velocities.
 
\subsection{Evolutionary models} \label{sec:evomodels}

We use the Galactic single star evolutionary models from \citet{brottRotatingMassiveMainsequence2011a}, henceforth referred to as the Brott grid. The grid has a mass range from \qtyrange{3}{100}{\msun} and contains models with initial equatorial rotational velocities from \qtyrange{0}{550}{\kms}. The models are calculated with the one-dimensional stellar evolution code described in \citet{hegerPresupernovaEvolutionRotating2000}, \citet{petrovicWhichMassiveStars2005}, and \citet{yoonEvolutionRapidlyRotating2005}. The initial mass fractions of hydrogen, helium, and metals are $X_i = 0.7274$, $Y_i = 0.2638$, and $Z_i = 0.0088$.

The major physics assumptions are as follows, with a more detailed discussion in \citet{brottRotatingMassiveMainsequence2011a}. Mixing and angular momentum transport are treated as diffusive processes. Angular momentum transport is handled with the Spruit-Tayler dynamo \citep{spruitDynamoActionDifferential2002}. Magnetic fields keep near rigid rotation while on the MS, which suppresses shear mixing between layers.

Convection uses the standard mixing length theory (MLT) as described by \citet{coxPrinciplesStellarStructure1968} and is modeled with the Ledoux criterion with a mixing length parameter $\alpha_\mathrm{MLT} = 1.5$. Semi-convection follows the prescription of \citet{langerSemiconvectiveDiffusionEnergy1983} and uses semi-convection parameter $\alpha_\mathrm{SEM} = 1$. Convective core overshooting uses a parameter of $0.335$ multiplied by the pressure scale height.
	
 Stellar wind mass loss is calculated with the prescriptions of \citet{vinkNewTheoreticalMassloss2000,vinkMasslossPredictionsStars2001}, which is based on the method described by \citet{koterEvolutionaryPhaseMass1997}, as well as the recipes from \citet{hamannSpectralAnalysesGalactic1995} and \citet{nieuwenhuijzenParametrizationStellarRates1990}. These two prescriptions predict strong mass loss at the bi-stability jumps found at $T_\mathrm{eff} \sim \qty{22}{\kk}$ and $T_\mathrm{eff}\sim \qty{12.5}{\kk}$. Notably, the mass loss rates in the Brott models are scaled directly with the adopted iron abundance, not with the overall metallicity, and are therefore not affected by adopted oxygen abundance. The Brott models adopt a rotationally induced mass loss enhancement, which becomes relevant for models near critical rotation. 
	
As discussed in Sect. \ref{sec:intro}, rapid rotation deforms stars and is difficult to model in a 1D code. We calculate the stellar structure equations and stellar properties on isobaric mass shells. This enforces the shellular approximation, which assumes that the angular velocity is constant on isobaric surfaces that correspond to mass shells. Centrifugal acceleration is accounted for with the methods of \citet{endalEvolutionRotatingStars1976} and \citet{kippenhahnSimpleMethodSolution1970}. The critical velocity is calculated as $v_\mathrm{crit} = \sqrt{\left( GM/r\right)\left(1-\Gamma_\mathrm{Edd}\right)}$, with $\Gamma_\mathrm{Edd} = L/L_\mathrm{Edd}$ as the Eddington factor and $L_\mathrm{Edd}$ as the Eddington luminosity.

We define the zero age main sequence (ZAMS) as the point where \qty{5}{\percent} of the initial core hydrogen mass fraction has been burned. For models with an initial mass of \qty{35}{\msun} and below, we define the terminal age main sequence (TAMS) as the beginning of the MS hook, which is the point at which core hydrogen has been exhausted and the stellar core contracts to raise the temperature in preparation for helium burning, which also raises the effective temperature ($T_\mathrm{eff}$).
Above this initial mass the models do not show a MS hook because stellar winds blow off the envelope before the end of the MS. For these models, we define TAMS as when the core hydrogen mass fraction drops to \num{0.03}.

We derive model predictions for the spin evolution in two ways. First, we use the original data from the Brott models. However, as a new method for calculating the the angular momentum loss has been recently devised (see Sect.\,\ref{amloss}), we develop a new prescription for the angular momentum loss, which is then used to derive a new prediction for the spin evolution.

\subsection{Angular momentum loss} \label{amloss}

In the Brott models, the amount of angular momentum lost due to stellar winds $\Delta J$ over one time step $\Delta t$ is computed by simply
removing the angular momentum contained in the outermost layer of
mass $\Delta M$ which is lost during the time step, i.e.,
by integrating the specific angular momentum profile, $j(m)$, as
\begin{equation} \label{eqn:og_angmo}
	\Delta J =J(t) - J(t-\Delta t) = \int^M_{M - \Delta M} j(m, t-\Delta  t) \:dm,
\end{equation}
where $m$ is the Lagrangian mass coordinate, and $\Delta M= \int^{t}_{t-\Delta t} \dot M\:dm$ for the specified stellar wind mass loss rate $\dot M$. In a rigidly rotating star, Eq.\,\ref{eqn:og_angmo} may lead to an underestimate of the angular momentum loss, which we can see as follows.

The time resolution of the evolutionary models is such that the stellar radius change over one time step $\Delta R$ is rather negligible, i.e., it is much smaller than $R-r(M-\Delta M)$, which is the difference between the stellar radius and the radius of the mass coordinate which will define the surface after the next time step, in a given stellar model. 

If we assume that the star remains in rigid rotation and neglect the change in stellar radius between two consecutive models, a more accurate way to compute the angular momentum loss is to consider a constant surface specific angular momentum $j_{\rm surf}$ and stellar mass loss rate over one time step \citep[c.f.,][]{paxtonModulesExperimentsStellar2019}, such that

\begin{equation}\label{eqn:new_angmo}
	\Delta J = \int_{t}^{t + \Delta t} j_{\rm surf}\:\dot{M} \,dt' =j_{\rm surf}\: \dot M \, \Delta t .
\end{equation}

\noindent This approach assumes that mass loss is independent of latitude and that $j_{\rm surf}$ is averaged over the latitude \citep{hegerPresupernovaEvolutionRotating2000}. 

\begin{figure}[!t]
	\centering
	\includegraphics{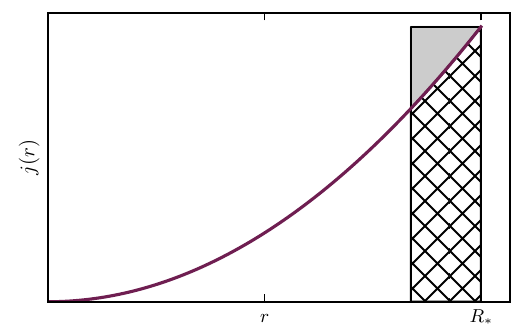}
	\caption{A schematic diagram of a model's specific angular momentum as a function of radius, illustrating the angular momentum loss prescriptions corresponding to Eqs.\,\ref{eqn:og_angmo} and \ref{eqn:new_angmo}. The width of the hatched rectangle indicates the radius range containing the material that is removed from the model within one time step. The prescription implemented in the Brott grid removes the angular momentum represented by the hatched area. The new prescription removes the angular momentum in the hatched plus the shaded area.}
	\label{fig:am_demo}
\end{figure}

Figure~\ref{fig:am_demo} illustrates the amount of angular momentum removed by both prescriptions. Notably, for small changes in mass per time step, the two prescriptions remove the same amount of angular momentum. If the radius of the bottom layer of the removed mass is not very close to the stellar radius, however, then the prescription originally implemented in the Brott grid results in the loss of a smaller amount of angular momentum, by an amount which depends on the mass loss rate and time steps used in the calculations.

While the angular momentum loss should not depend on the time step of the numerical calculations, the difference between the old and our new prescription may reflect the magnitude of the uncertainty of the angular momentum loss predicted by our 1D stellar evolution models. Our new approach may be more correct, but it depends on the assumption that the stellar outer layers remain essentially rigidly rotating. 

The evolution of stars is essentially independent of their initial rotation velocity for the range of rotation rates in this study, and thus it is simple and accurate to apply the new prescription for angular momentum loss to the Brott models.
We used Eq.\,\ref{eqn:new_angmo} to recalculate the rate of angular momentum loss in the Brott grid and used the results to compute revised values for $J$, $j_{\rm surf}$, and $v_{\rm rot}$.

\subsection{Population synthesis} \label{sec:popsynthalgo}

We developed the population synthesis algorithm H\=ok\=u (\citealt{nathanielSpindownEnvelopeInflation2022}, available on request) in order to study the statistical properties of MS single stars. Using the appropriate distribution functions, H\=ok\=u takes a specified initial mass ($m_\mathrm{i}$), ZAMS rotational velocity ($v_\mathrm{ZAMS}$), and age ($\tau_*$) and selects four model stars from the grid. The algorithm then interpolates stellar parameters from these four tracks, using a three-dimensional linear interpolation algorithm. This process is repeated many times until H\=ok\=u has interpolated a population of sufficient size.
	
H\=ok\=u interpolates a simulated star as follows: after drawing $m_\mathrm{i}$, $v_{\rm ZAMS}$, and $\tau_*$, it selects the two closest initial masses above and below $m_\mathrm{i}$ in the model grid. For each of these masses, H\=ok\=u then selects the two model stars with ZAMS velocities above and below $v_{\rm ZAMS}$. These are the four models used in interpolation. A model's rotational velocity can change between the beginning of the simulation and ZAMS, especially for massive stars, which is why H\=ok\=u uses the models' ZAMS velocity, rather than their initial velocity.
		
The MS lifetime ($\tau_\mathrm{MS}$) depends on both initial mass and rotation rate, so the four models are not necessarily in the same evolutionary stage at the same age. To combat this, we use the fractional MS age, $f_\mathrm{MS} = \tau_*/\tau_\mathrm{MS}$, as the interpolation variable, rather than just $\tau_\mathrm{MS}$. H\=ok\=u interpolates an MS lifetime for the simulated star from the MS lifetimes of the four model stars and uses this to determine the simulated star's $f_\mathrm{MS}$. If the simulated star's $f_\mathrm{MS} > 1$, however, it is thrown out and H\=ok\=u moves to the next initial parameter set. This ensures that the population consists of only MS simulated stars. 
		
\begin{figure}
				\centering
				\includegraphics{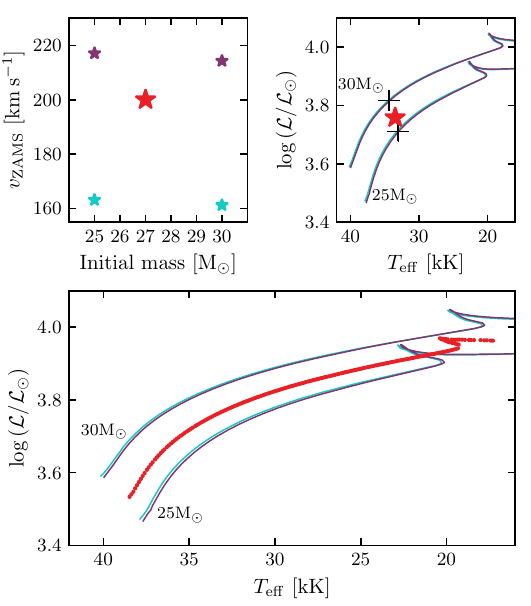}
				\caption{\textit{Top}: Illustration of the H\=ok\=u interpolation process. Left panel shows the initial masses and ZAMS velocities of the grid models in purple and turquoise and the simulated star in red. The right panel shows the evolutionary models in a spectroscopic HR diagram and the simulated star's interpolated effective temperature and spectroscopic luminosity. The crosses mark where the effective temperature and spectroscopic luminosity are interpolated from the grid models. \textit{Bottom}: the resulting evolutionary track of the \qty{27}{\msun}, $v_{\rm ZAMS} = \qty{200}{\kms}$ interpolated star in a spectroscopic HR diagram, along with the grid models used for interpolation.}
				\label{fig:interp_demo}%
			\end{figure}

Once this process is complete, H\=ok\=u can interpolate observables for the simulated star from the four stellar models. In addition to population synthesis, H\=ok\=u can also calculate isochrones and evolutionary tracks. A visual representation of the interpolation process and an interpolated evolutionary track is shown in Fig.\,\ref{fig:interp_demo}.
		
\section{Model predictions} \label{sec:model_predictions}
	
		\begin{figure*}[!htbp]
		\centering
		\includegraphics{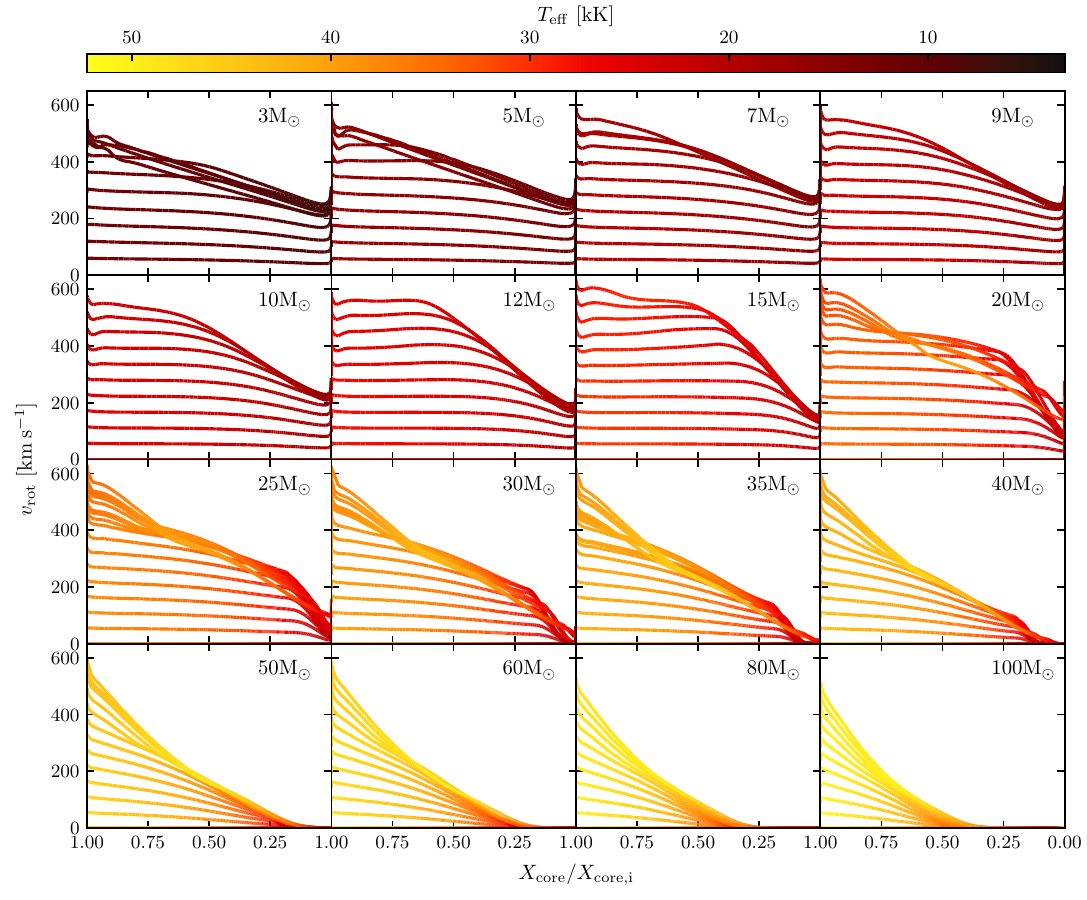}
		\caption{Evolution of the rotational velocity from ZAMS to TAMS of the models from the Brott grid, according to the new predictions (derived as discussed in Appendix \ref{app:AM_loss}). Each panel shows a different initial mass with the colors indicating the effective temperature over the course of the MS. The $x$-axis shows the current center hydrogen mass fraction ($X_\mathrm{core}$) divided by the initial center hydrogen mass fraction ($X_\mathrm{core,i}$).}
		\label{fig:brott_spindown}
	\end{figure*}

Figure \ref{fig:brott_spindown} shows the MS evolution of the surface rotation velocity of the Brott models from \qtyrange{3}{100}{\msun} based on our new predictions. We see a convergence of the lines with time (or with burnt hydrogen fraction) in all panels. This occurs mainly because the rate of angular momentum loss is proportional to the surface specific angular momentum, which implies that faster rotators spin down faster than slower rotators. On top of this feature, there is a distinct dependence of the spin down on the initial mass of the stars, which is due to the mass dependence of the impact of the three processes affecting the spin evolution mentioned in Sect.\,\ref{sec:intro}: expansion, internal transport, and mass loss.

In stars with masses below 10\,\msun, the core masses are small, and they contain little angular momentum. Therefore, the core contraction during hydrogen burning and the corresponding angular momentum transport from the core into the expanding envelope has little effect, and expansion leads to spindown of the star.

In the mass range \qtyrange{10}{20}{\msun}, the core mass fractions are larger. Their contraction triggers an angular momentum flow from the core into the envelope, which contributes significantly to counteract the spindown of the surface layers (cf.\,\citealt{hastingsSingleStarPath2020} for an in-depth discussion). In particular the 12$\msun$ and 15$\msun$ models even show a spin-up during the first part of their hydrogen burning evolution. Since stars are rarely born rotating extremely rapidly \citep{bodenheimerAngularMomentumEvolution1995}, the majority of stars in this mass range are predicted to evolve with nearly constant rotational velocity over most of their hydrogen burning lifetime. However, the mass-loss effect also kicks in in this mass range, especially during the second half of the MS evolution, during which the stars are on the cool side of the wind bi-stability regime where the mass loss rates are significantly larger \citep[cf.][although some new evidence may refute this, see e.g., \citealt{deburgosIACOBProjectXI2024}]{vinkNatureSupergiantsClues2010}.

For stars above \qty{\sim 25}{\msun}, mass loss dominates the spin evolution from the begining, and the mass loss induced spindown is so strong that all models end core hydrogen burning with surface rotation velocities near zero. As the effect becomes stronger with mass, the most massive stars considered here, with a mass above \qty{\sim 50}{\msun}, reach this situation during core hydrogen burning, and live a significant fraction of it essentially without rotation. This effect is augmented by the envelope inflation of the models near their Eddington limit for masses above \qty{\sim 30}{\msun}. \citep{sanyalMassiveMainsequenceStars2015, sanyalMetallicityDependenceEnvelope2017}

Although in Fig.\,\ref{fig:brott_spindown} we only present results for the spin rate evolution with our new prescription for angular momentum loss, it is interesting to briefly mention the main impacts of considering either prescription. In App.\,\ref{app:AM_loss}, we present a comparison between the two prescriptions for the total angular momentum, surface specific angular momentum, and rotational velocity quantities.

Below \qty{\sim 10}{\msun}, the mass loss effect is small, and our new results are very close to the original prediction of the Brott models. Even at \qty{20}{\msun}, the difference for a model rotating initially with about \qty{200}{\kms} remains within a few percent (cf. Fig.\,\ref{fig:am_corrections_20msun} in App.\,\ref{app:AM_loss}). However, for higher masses, the difference becomes quickly more appreciable. Considering stars rotating initially with about \qty{200}{\kms}, we find they spin down completely, while the original Brott model spins down to about \qty{80}{\kms}. The difference between the original and our new calculation is largest at the top end of the considered mass spectrum. In the original calculations, the \qty{100}{\msun} model starting with \qty{\sim 200}{\kms} spins down to \qty{135}{\kms} when half of the hydrogen is burnt in the center, while it reaches \qty{\sim 50}{\kms} with our new approach at the same moment in time. Complete spin down ($v_{\rm rot} < \qty{10}{\kms}$) is reached at a core hydrogen mass fraction of 0.14 compared to 0.30, respectively.

\section{Observational sample and preprocessing} \label{sec:obs_sample}
	
The observations used in this work come from a subsample of the stars for which the IACOB project \citep{simon-diazIACOBProjectSynergies2011,simon-diazIACOBSpectroscopicDatabase2020} have compiled high-quality spectroscopic observations. In particular, we focus in those samples of O stars and B-type supergiants described and spectroscopically analyzed by \citet{holgadoIACOBProjectVI2020, holgadoIACOBProjectVII2022} and \citet{deburgosIACOBProjectIX2023, deburgosIACOBProjectX2024}, respectively, reaching a total of about 800 Galactic blue massive stars.
	
As described in the above mentioned papers, the considered samples exclude all stars classified as double-line spectroscopic binaries (SB2) and only concentrates in stars identified as likely single stars and single-line spectroscopic binaries (SB1). 
While eliminating the SB2 seems appropriate for comparison to single star population synthesis, we remark that, following \citet{deminkRotationRatesMassive2013}, 
a non-negligible percentage of the remaining sample of stars categorized as apparently single and SB1 might be the outcome of a binary interaction event.

Throughout the paper, we will be using the spectroscopic Hertzsprung-Russell diagram (sHRD, $\mathscr{L} \equiv T_\mathrm{eff}^4/g$ with $g$ as surface gravity; \citealt{langerSpectroscopicHertzsprungRussellDiagram2014}) to compare our results with the observed stars, which eliminates uncertainties in reddening and distance.

	\subsection{Kernel density estimate}\label{sec:kde_estimate}

We compare the observed stars to the synthetic populations by constructing a two-dimensional kernel density estimate (KDE) for each observed star in the sample, placed in the spectroscopic Hertzsprung-Russell diagram, and normalizing it so that the sum is unity (obtaining the full sample KDE is simply a matter of adding the individual KDEs together). The uncertainties in $T_{\rm eff}$ and $\mathscr{L}$ are used as the standard deviations in the KDEs. Integrating an individual star's KDE over an arbitrary area on the sHRD gives the probability that the star is located in that area and therefore the weight it should be given in constructing a distribution of the stellar parameter in question. Accordingly, every star is considered when constructing distributions, but the magnitude of any given star's contribution (the "weighting factor") will vary based on its location relative to the chosen area on the sHRD. 
 
\begin{figure}[!htbp]
		\centering
		\includegraphics{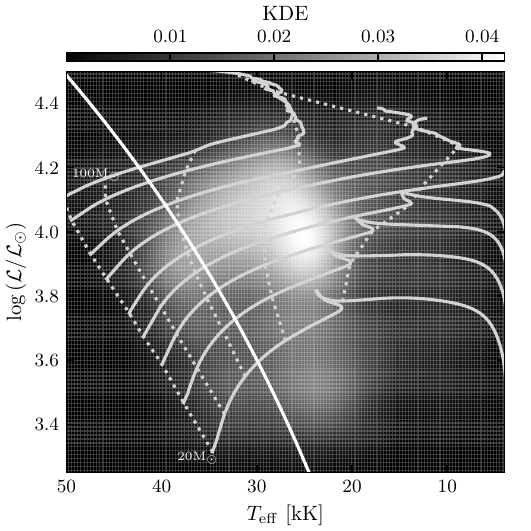}
		\caption{The background represents the two-dimensional KDE of the entire observed sample in the sHRD. The solid lines show the Brott grid evolutionary tracks at \qty{20}{\msun}, \qty{25}{\msun}, \qty{30}{\msun}, \qty{35}{\msun}, \qty{40}{\msun}, \qty{50}{\msun}, \qty{60}{\msun}, \qty{80}{\msun}, and \qty{100}{\msun}. The dotted lines divide the MS into quarters, determined by the fraction of the initial center hydrogen mass fraction that has been burnt. The white line indicates $\log g = \qty{3.7}{dex}$. Only stars to the left of this line are used to create the initial velocity KDE discussed in Section \ref{sec:vini_pdf}.
  }
		\label{fig:2dkde}
	\end{figure}

 Summing these weighting factors results in a number, not necessarily an integer, to which the data can be considered equivalent. E.g., a distribution that sums to $10$ could consist of data from $5$ stars that each have a weighting factor of \qty{20}{\percent}, or $100$ stars that each have a weighting factor of \qty{10}{\percent}. This method allows us to construct distributions for selected portions of the sHRD while taking observational uncertainties into account. We do not introduce any smoothing and conversely do not attempt to deconvolve our observational errors. The result can be seen in Fig.\,\ref{fig:2dkde}.

\subsection{Initial rotation velocity PDF} \label{sec:vini_pdf}

We chose to use the observed sample to construct an initial velocity distribution for use in our population synthesis algorithm. In order to exclude more evolved stars, we only used stars with $\log g > 3.7$\,dex in constructing our distribution (this line is marked in Fig.\,\ref{fig:2dkde}). Notably, according to the stellar models, this condition selects stars during the first half of their MS evolution, while a higher threshold surface gravity would select stars closer to the ZAMS. However, the number of sample stars drops quickly towards the ZAMS \citep{holgadoIACOBProjectVI2020}, such that a higher threshold gravity would not give us enough statistics. It implies that we can only constrain the spin down evolution of Galactic massive stars in the second half of their MS evolution.
	
We explore two different fit methods for the data, one in linear space and one in logarithmic space. For the linear space fit we used the \verb*|scikit-learn| GaussianMixture method \citep{pedregosaScikitlearnMachineLearning2011} for a 2-component Gaussian. This fit accurately reproduces the feature seen at \qty{\sim 200}{\kms} in the histogram, but allows for negative velocities in the PDE. For the logarithmic space fit we first took the logarithm of the data and then fit it with the \verb*|scipy.stats| \citep{virtanenSciPy10Fundamental2020} 1-component Gaussian KDE function. This logarithmic space fit was not high enough resolution to capture the feature at \qty{\sim 200}{\kms}, but it does not allow for unphysical negative velocities.

\begin{figure}[!htbp]
		\centering
		\includegraphics{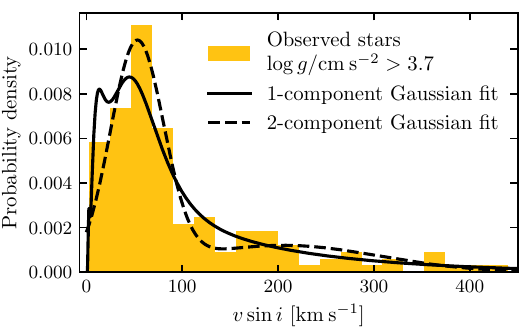}
		\caption{Distribution of projected rotational velocity for observed stars with $\log g > \qty{3.7}{dex}$, overplotted with 1- and 2-component Gaussian KDEs (solid and dashed lines, respectively).}
		\label{fig:vsini_fit}
	\end{figure}

We find that the observed sample's $v\sin i$ distribution is not well characterized by a single Gaussian in linear or log space and there is an overabundance in density at larger velocities relative to a single Gaussian fit.
We performed a linear regression with \verb*|scipy.stats| and found no statistically significant differences in their performance.
Therefore, we chose to use the KDE fitted in log space for H\=ok\=u, as it does not allow negative velocities and is simpler to work with.
The histogram and the two fits are shown in Fig.\,\ref{fig:vsini_fit}.
	
	\section{Comparison with observations} \label{sec:compare_to_obs}

    \begin{figure*}[!htbp]
		\centering
		\includegraphics{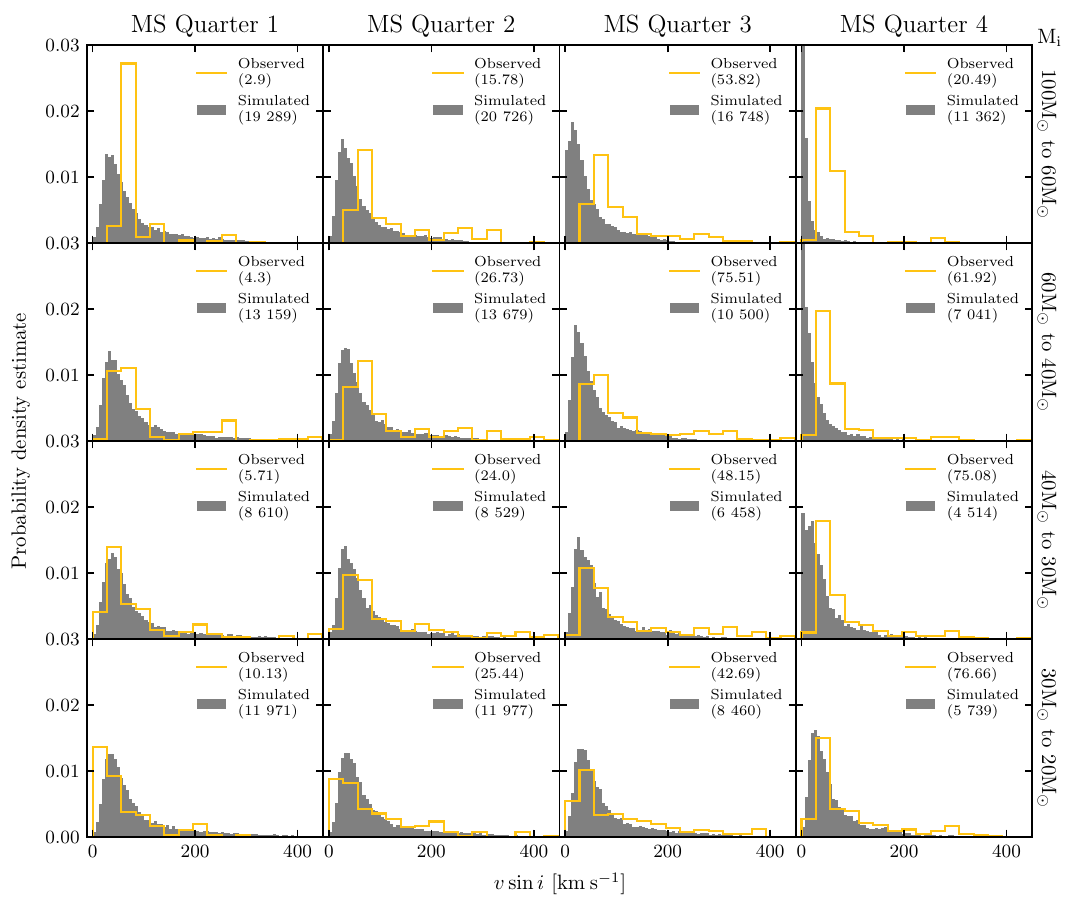}
		\caption{Distributions of projected rotational velocity, $v\sin i$ for the observed stars (yellow lines) and for the synthetic population (gray histograms), where the $v\sin i$ is calculated using the rotational velocities directly from the models of \citet{brottRotatingMassiveMainsequence2011a}. The four columns show the different quarters of the MS, as denoted above, and the four rows show the different mass bins, as denoted to the right. The numbers in the legends indicate the number of observed or simulated stars in the respective histograms (cf. Sect.\,\ref{sec:kde_estimate}). The simulated distribution reaches beyond the $y$-axis boundary in two panels, where for the \qtyrange{100}{60}{\msun} mass bin, the fourth quarter reaches to $0.13$, and for the \qtyrange{60}{40}{\msun} mass bin, the fourth quarter reaches to $0.069$. }
		\label{fig:iacob_spindown_uncorr}
	\end{figure*}

		\begin{figure*}[!htbp]
		\centering
		\includegraphics{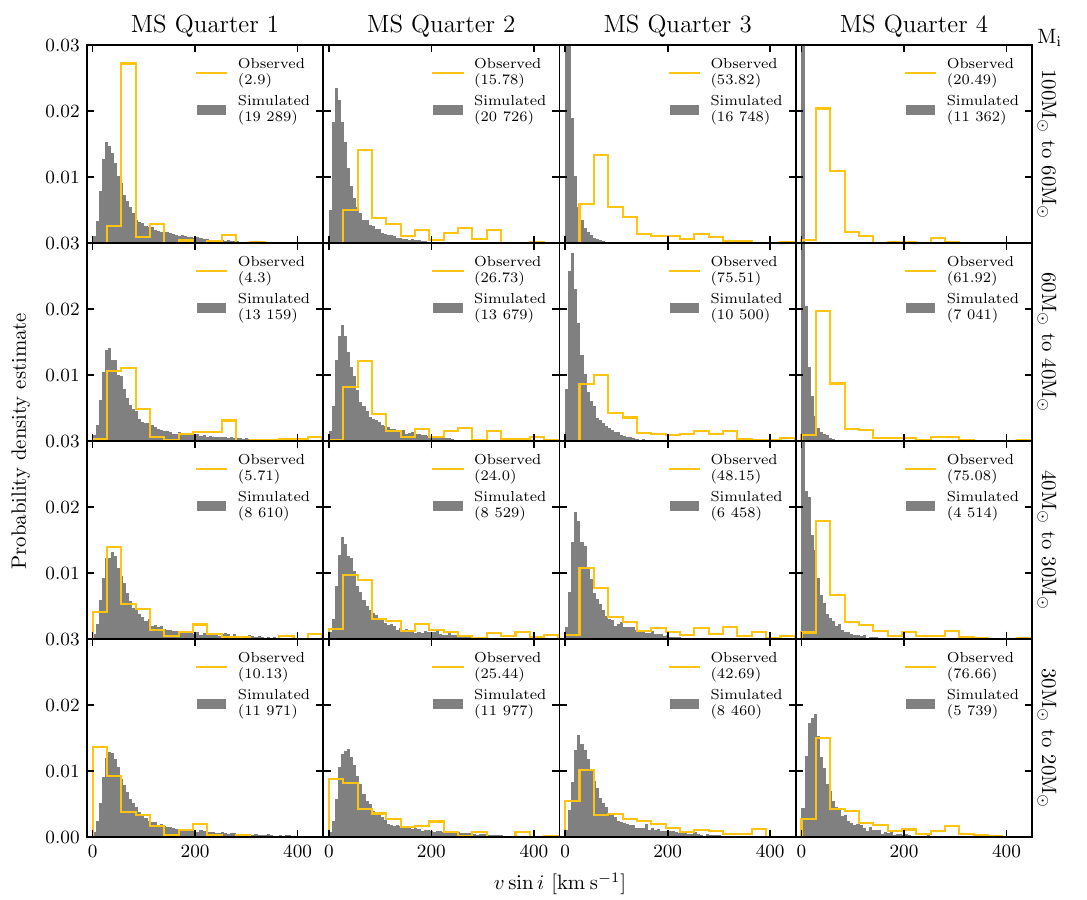}
		\caption{As Fig.\,\ref{fig:iacob_spindown_uncorr}, but with $v\sin i$ values for the synthetic population calculated by applying our new angular momentum loss recipe to the models of \citet{brottRotatingMassiveMainsequence2011a}, as discussed in Appendix \ref{app:AM_loss}. The simulated distribution reaches beyond the $y$-axis boundary in four panels. For the \qtyrange{60}{100}{\msun} mass bin, the third quarter reaches to $0.08$ and the fourth quarter reaches to $0.17$. For the \qtyrange{40}{60}{\msun} mass bin, the fourth quarter reaches to $0.12$. For the \qtyrange{30}{40}{\msun} mass bin the fourth quarter reaches to $0.05$. }
		\label{fig:iacob_spindown}
	\end{figure*}

In this section, we compare the observed $v\sin i$-distributions with our theoretical predictions using H\=ok\=u.
Here, we first consider the results taken directly from the models of Brott et al. and compare them with the IACOB data. In a second step, we compare the observational data with the $v\sin i$ distributions derived from the Brott models after applying our new angular momentum loss prescription.

We use a flat initial mass function (IMF) from \qtyrange{20}{100}{\msun} and assume continuous star formation. We draw inclination angles ($i$) from a PDF that assumes random orientation $\cos i$. We use the distribution function discussed in Sect.\,\ref{sec:vini_pdf} to draw a $v_{\rm ZAMS} \sin i$ for each simulated star and then divide by $\sin i$ to obtain $v_{\rm ZAMS}$, which ranges from \qtyrange{\sim 1}{\sim 560}{\kms}		
		
We first bin the synthetic population by initial mass, using 4 bins of \qtyrange{20}{30}{\msun}, \qtyrange{30}{40}{\msun}, \qtyrange{40}{60}{\msun}, and \qtyrange{60}{100}{\msun}. We choose the first \qtyrange{20}{30}{\msun} mass bin because winds are less significant at this range than those above \citep{vinkTheoryDiagnosticsHot2022}. The other mass bins are chosen to be approximately evenly spaced in $\log \mathscr{L}/\mathscr{L}_\odot$. 
		
To facilitate seeing how rotational velocity changes with time, we divide the MS into quarters based on how much of the initial core hydrogen has been burnt. E.g., the first quarter consists of simulated stars that have a core hydrogen mass fraction of $X_i$ to $0.75 X_i$ and the last quarter consists of simulated stars with a core hydrogen mass fraction of $0.25 X_i$ to TAMS. This technique allows us to see the mass-dependent evolution of rotational velocity across the MS.

\subsection{Comparing with the original Brott model predictions}

Focusing first on the theoretical data, Fig.\,\ref{fig:iacob_spindown_uncorr} shows an overall spindown of stars with time, in all four considered mass bins. In the lowest mass bin (\qtyrange{20}{30}{\msun}), the spindown is mild, barely shifting the peak of the distribution to smaller $v \sin i$-values. Only the initially fastest rotators are moved substantially, such that barely any stars with $v \sin i > 250\,$km\,s$^{-1}$ is predicted. For higher masses, the predicted spindown gets markedly stronger, with upper limits to the $v \sin i$ values of $210\,$km\,s$^{-1}$, $180\,$km\,s$^{-1}$, and $100\,$km\,s$^{-1}$, in the last quarter of the three higher mass bins. 

The spindown in the models of all four mass bins accelerates with time. The model histograms in the first two quarters are almost identical, even in the highest mass bin. For the lower two mass bins, even the predicted distributions in the second and third quarter are nearly the same. The biggest differences in subsequent quarters occurs between the third and the fourth, still rather modest in the lowest mass bin (\qtyrange{20}{30}{\msun}).

When we compare the predictions with the observed $v \sin i$-distributions, starting with the lowest mass bin, we find rather good agreement. While this is unsurprising for the first two quarters, which contains the stars used for designing our initial $v \sin i$-distribution, it confirms that angular momentum transport from the core into the envelope of the stars must take place. The stellar radii increase by a factor of $\sim 2$ from the second to the fourth quarter, such that local angular momentum conservation would lead to a spindown by the same factor, which is neither observed nor predicted. On the other hand, we see that the fastest rotating models ($v \sin i > 200\,$km\,s$^{-1}$) are spun-down in the last quarter, while the number of observed fast rotators is increasing. While the total number of such stars is small, we see this effect in all four mass ranges, which indicates that it is significant; we come back to this below.

In the \qtyrange{30}{40}{\msun} mass range, the situation is similar to the \qtyrange{20}{30}{\msun} range. Here, and in the higher mass ranges, the number of observed stars in the first quarter is small and the corresponding histogram is not overly meaningful. In the last two quarters, the spindown in both the observed and the model distributions is similar to the lower mass range, perhaps slightly more pronounced in the final quarter. Most notably in the fourth quarter, however, the predicted distribution is more strongly skewed towards $v \sin i =0$ than the observed distribution. This could imply that the angular momentum loss in the models is overestimated. On the other hand, the observed stars in the fourth quarter are giants or supergiants, which are known to be strongly affected by additional non-rotational sources of line-broadening. In particular, both macroturbulent and microturbulent broadening might play a significant role in the reliability of the $v\sin i$ estimates (below a given threshold) if the relative contribution of the three broadening mechanisms is not adequately disentangled in the analysis of the line-profile \citep[see, e.g., detailed explanations and illustrative examples in][]{simon-diazIACOBProjectRotational2014, simon-diazIACOBProjectIII2017, holgadoIACOBProjectVII2022}. As indicated in \citet{holgadoIACOBProjectVII2022} and \citet{deburgosIACOBProjectIX2023}, our $v\sin i$ measurements are properly decontaminated from macroturbulent broadening; however, it is possible that microturbulent broadening imposes a limit on the line-broadening analysis that our state-of-the-art methodologies perform to reliably reach values of $v\sin i$ below \qtyrange{20}{40}{\kms} in certain parameter domains. Therefore, the reason of the mismatch between the predicted and observed number of slow rotators in the fourth quarter of the \qtyrange{30}{40}{\msun} mass range could be also associated with limitations in our availability to obtain reliable of projected rotational velocities in the low $v\sin i$ domain, as further discussed in \citet[][in particular, their Sect.\,3.2.3 and App.\,D]{holgadoIACOBProjectVII2022}.

For the top two mass bins in the figure (\qtyrange{100}{60}{\msun} and \qtyrange{60}{40}{\msun}), ignoring the first quarter, the observed distributions do not change strongly with evolutionary time. In particular, the distributions of the second and third MS quarter are very similar. This seems to imply that, even at the highest considered masses, the rotational velocities are not changing significantly during the first \qty{75}{\percent} of the MS evolution. This is remarkable because especially in the third quarter, the models predict a significant spindown. The discrepancy is strongest in the fourth quarter, for which the observations signify some spindown, but by far not as much as predicted. Whether this rather large discrepancy can be masked by observational limitations (imposed by macro- or micro-turbulence) when measuring projected rotational velocities appears doubtful.

\subsection{Comparing with the updated Brott model predictions}

We show the comparison of the IACOB data with the distributions of $v\sin i$ calculated with our new angular momentum loss approach in Fig.\,\ref{fig:iacob_spindown}, which is structured in the same way as in Fig.\,\ref{fig:iacob_spindown_uncorr}. Since the new predictions of all four panels of the lowest mass range, and of the first three panels of the \qtyrange{30}{40}{\msun} range are almost identical with those from the original Brott model data, and even in the 4th quarter of the latter mass range the changes are small, we only discuss the two top mass ranges here.

The angular momentum loss according to our new prescription is larger than that in the original Brott models. Therefore, in particular in the \qtyrange{40}{60}{\msun} and \qtyrange{60}{100}{\msun} mass ranges, the predicted spindown is stronger. As the spindown of the very massive original Brott models appeared stronger than that implied by the IACOB data, this discrepancy becomes worse when comparing to the new predictions. Already in the third MS quarter, the models cover less than half of the observed distributions in both mass ranges. In the fourth quarter, the models cover \qty{\sim 5}{\percent} of the observed distribution for \qtyrange{40}{60}{\msun}, and the overlap between observed and predicted distributions is essentially zero in the highest mass bin. We conclude that our new angular momentum loss prescription overpredicts the wind induced angular momentum loss in stellar models.

\section{Discussion} \label{sec:discuss}

From the comparisons of the model predictions with the IACOB data, we can draw the following conclusions. It appears that for Galactic stars below \qty{\sim 40}{\msun}, the available predictions are not sensitive to the theoretical uncertainties. The main feature of the models is that for rotational velocities below \qty{\sim 200}{\kms}, which covers $\simgr\qty{ 90}{\percent}$ of the observed stars, the rotational velocity remains rather constant throughout the MS evolution. This trend is in good agreement with the observed rotational velocity distributions. It implies that the rotation of the expanding envelope is strongly coupled to the rotation of the contracting core. 

For stars of higher mass ($\simgr \qty{40}{\msun}$), signs of angular momentum loss are clearly visible in the observed $v \sin i$ distributions, as the peak of the distribution moves to slower $v\sin i$ values as the stars progress across the MS. The data provides a vivid test of the adopted angular momentum loss prescriptions in stellar evolution models. While on grounds of physical consistency, our new prescription appears preferable over the one used in the original Brott models, it clearly overpredicts the angular momentum loss when applied to the Brott model data. While the original Brott models also lead to slower rotation than what is derived from the IACOB data, this difference could also be related to uncertainties in relation to atmospheric micro-turbulence in the evolved very massive stars.

Independent of the used angular momentum loss prescription, the models predict that any initially very fast rotator ($v_{\rm rot} \simgr \qty{200}{\kms}$) should have been spun down by its wind to lower velocities by the end of the MS evolution.

The continued presence of very fast rotators in samples of more and more evolved massive stars (MS quarters 2, 3, and 4 in Figs.\,\ref{fig:iacob_spindown_uncorr} and\,\ref{fig:iacob_spindown}) implies that these are unlikely to be evolved stars which were born very rapidly rotating, but instead that they have been spun-up at one point during their MS evolution, as predicted by models of mass-transferring binary stars \citep{langerPresupernovaEvolutionMassive2012,deminkRotationRatesMassive2013}.

One way to alleviate the problem of a strong spindown for stellar models above \qty{\sim 40}{\msun} for a given angular momentum loss prescription, would be to lower the adopted mass loss rates. The Brott models use \citet{vinkMasslossPredictionsStars2001}, which has been criticized for an unduly high mass loss rate \citep[e.g.,][]{bjorklundNewPredictionsRadiationdriven2023}. However, the Vink prescription has been found to accurately reproduce the rates of hot OB stars recently measured with high quality UV and optical spectra \citep{verhammeXShootingULLYSESMassive2024}.

Another way to reduce the mass loss rate would be to reconsider the inclusion of the so called bi-stability jump, which is a rapid increase in mass loss near $T_{\rm eff} = \qty{22000}{\kelvin}$ with corresponding strong spindown \citep{vinkNatureSupergiantsClues2010} that is also seen in the Brott models. The bi-stability jump has been brought into question by recent observations \citep{verhammeXShootingULLYSESMassive2024, deburgosIACOBProjectXI2024}, implying that the predicted spindown may not exist.

Fig.\,\ref{fig:2dkde} shows that the bi-stability breaking is expected to affect the predictions in the fourth MS quarter, where the discrepancy between models and observations is indeed the largest (cf. Figs.\,\ref{fig:iacob_spindown_uncorr} and\,\ref{fig:iacob_spindown}). These two figures also show, however, that the spindown problem already occurs in the third MS quarter, strongest in Fig.\,\ref{fig:iacob_spindown} based on results using the new angular momentum loss scheme. Since faster rotators spin down faster than slower rotators (cf. Sect.\,\ref{sec:model_predictions}), it seems unlikely that a general downwards correction of the mass loss rates by a factor of $\sim 2$ for stars between \qty{40}{\msun} and \qty{100}{\msun} would be sufficient to resolve the issue, as the rotational velocities will come down by a smaller factor. Thus, revised mass loss rates could resolve the tension, but the revision would need to be substantial.

Strong model uncertainties also originate from the proximity of the massive stars to their Eddington limit. First, it leads to vigorous subsurface convection zones in these stars \citep{cantielloSubsurfaceConvectionZones2009}, which has been suggested to lead to atmospheric macroturbulence \citep{grassitelliRelatingTurbulentPressure2015, grassitelliMetallicityDependenceTurbulent2016}. As the convective energy transport in the non-adiabatic part of these zones is inefficient, these layers may inflate and increase the stellar radius \citep{sanyalMassiveMainsequenceStars2015, sanyalMetallicityDependenceEnvelope2017}. We show in App.\,\ref{app:inflation} that this radius increase is sensitive to the chosen mixing length parameter $\alpha_\mathrm{MLT}$. For Galactic stars above \qty{\sim 40}{\msun}, the speed at which they increase their radius, as well as the fractions of their lifetime spent in the hotter and cooler part of the HRD is depending on this parameter. Furthermore, \citet{vinkWindModellingVery2011} found a sharp increase in the slope of the mass loss rate vs Eddington limit relation, from $\dot{M}\propto \Gamma^2$ for stars with $\Gamma \lesssim 0.7$ to $\dot{M} \propto \Gamma ^5$ for $\Gamma > 0.7$. Thus, it is possible that stars near the Eddington limit have a more rapid spindown, while stars further from the Eddington limit might be spinning down more slowly.

We note that we only consider single star evolution here, despite the fact that over \qty{70}{\percent} of all massive O-type stars will interact with a companion \citep{sanaBinaryInteractionDominates2012}. However, \citet{deminkIncidenceStellarMergers2014} showed that only about \qty{10}{\percent} of the massive MS stars in a stellar population are expected to be merger products or accretion stars, respectively. While those could be related to the tail of rapid rotators found in the observed rotational velocity distributions (Figs.\,\ref{fig:iacob_spindown_uncorr} and\,\ref{fig:iacob_spindown}), they are unlikely to shape their main peaks. Finally, we note that there are several additional uncertainties associated with the evolution of massive stars close to critical rotation \citep[e.g., see][]{hastingsModelAnisotropicWinds2023, petrenz2nonLTEModelsRadiation2000, bogovalovWindsFastRotating2021}, which we do not address here in detail since the vast majority of the observed stars are not close to this situation.

	\section{Conclusions} \label{sec:conclusions}

In this paper, we compare the rotation velocities of stars in a synthetic populations based on single star evolutionary models to those of an observed sample of \num{\sim 800} Galactic OB stars. We find general agreement for stars below \qty{\sim 40}{\msun}, which implies that the rotation of the stellar envelope is well coupled to that of the stellar core. However, our models of more massive stars spin down faster than what is implied from the observed stars (cf. Figs.\,\ref{fig:iacob_spindown_uncorr} and\,\ref{fig:iacob_spindown}). We have limitations for very low $v\sin i$ measurements, and that affects the comparison with very strong spindown.

We find this result when applying two different angular momentum loss prescriptions, where the discrepancy is in fact larger when we use our new method, which is based on more consistent physics assumptions. As described by \citet{holgadoIACOBProjectVII2022}, a similar result was found when using the stellar models of \citet{ekstromGridsStellarModels2012}. As discussed in Sect.\,\ref{sec:discuss}, substantially lower stellar wind mass loss rates could resolve this tension, but currently suggested rates are unlikely to resolve it. 

Our results imply further that we are unable to provide significant constraints on the initial rotation rates of stars above \qty{\sim 20}{\msun}. This is so due to the scarcity of observed massive stars during the first half of their life \citep[see also][]{holgadoIACOBProjectVI2020}, combined with the model prediction that even initially fast rotators would spin rather slowly during the second half of their MS evolution (cf. Fig.\,\ref{fig:brott_spindown}). The continued presence of a tail of fast rotators in the observed rotational velocity distributions for evolved MS stars appears to require a spin-up of a small fraction of the stars at one point during their MS evolution, e.g., by mass transfer from a companion star.

	\bibliographystyle{aa} 
	\bibliography{refs} 

\begin{appendix} 
	\section{Example models at $20\mso$ and $40\mso$}\label{app:AM_loss}
	
		        \begin{figure}[!b]
						\centering
						\includegraphics{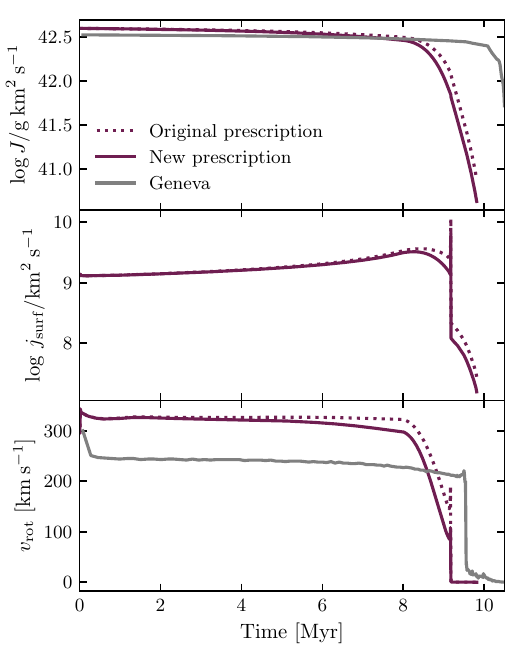}
						\caption{Evolution of total angular momentum (top), surface specific angular momentum (middle), and rotational velocity (bottom) for a \qty{20}{\msun} Brott grid model with $v_\mathrm{i} = \qty{300}{\kms}$, which is $\sim\,0.4 \times v_\mathrm{crit}$. The dotted line indicates the original calculation, the solid line shows the value recalculated using the new angular momentum loss prescription. A \qty{20}{\msun} Geneva model \citep{eggenbergerGridsStellarModels2021} is shown in gray in the top and bottom panels, with initial rotation set to $0.4\times v_\mathrm{crit}$.}
						\label{fig:am_corrections_20msun}
					\end{figure}
					
				    \begin{figure}[!b]
						\centering
						\includegraphics{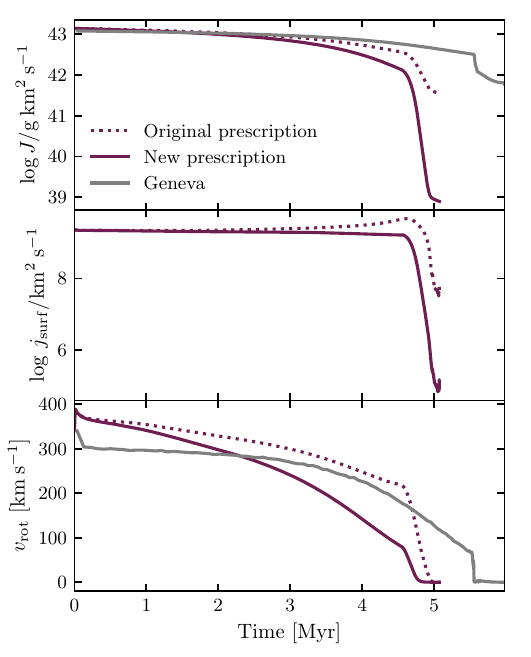}
						\caption{In the same style as Fig.\,\ref{fig:am_corrections_20msun}, but for a \qty{40}{\msun} Brott grid model with $v_\mathrm{i} = \qty{350}{\kms}$, which is $\sim0.4 \times v_\mathrm{crit}$. The Geneva model \citep{eggenbergerGridsStellarModels2021} shown in the top and bottom panels is \qty{40}{\msun} with rotation set to $0.4\times v_\mathrm{crit}$.}
						\label{fig:am_corrections_50msun}
					\end{figure}

We show the result of applying the new angular momentum loss prescription to a Brott grid model in Figs.\,\ref{fig:am_corrections_20msun} and \ref{fig:am_corrections_50msun}, as well as a comparison to models calculated with the Geneva stellar evolutionary code.

We use the one-dimensional models of \citet{eggenbergerGridsStellarModels2021} as a comparison to our Brott grid models. These models are calculated with GENEC, the Geneva stellar evolutionary code. They have an initial metallicity of $Z = 0.006$, i.e., near a star in the Large Magellanic Cloud. These models take a different approach to angular momentum transport than \citet{brottRotatingMassiveMainsequence2011a}, accounting for hydrodynamical transport by the shear instability and meridional currents, and not considering internal magnetic fields. Therefore, they experience significant differential rotation during the MS, which is not seen in the \citet{brottRotatingMassiveMainsequence2011a} models.

Fig.\,\ref{fig:am_corrections_20msun} shows only small differences between the original and new data, as lower mass stars are less affected by stellar winds. The differences become more pronounced as the star approaches the end of its life, but are still very small. The Geneva model lives longer and does not show the sharp increase in angular momentum loss towards the end of its life. While spindown increases rapidly at \qty{\sim 9.8}{\myr} there is only a small increase in angular momentum loss rate at the same time.

In Fig.\,\ref{fig:am_corrections_50msun} there is a steep decrease in angular momentum at \qty{\sim 4.75}{\myr}, which corresponds to an increase in spindown and loss of surface specific angular momentum. The model reduces its rotational velocity at a more linear rate with our new angular momentum loss scheme, as opposed to the original approach, which shows two clear phases, before and after the \qty{\sim 4.75}{\myr} mark. There is still a small increase in spindown with the new approach at that point, but since the model has already spun down so much it has minimal influence. Overall, the new scheme increases the amount of angular momentum lost, thereby reducing the impact of angular momentum transport and smoothing the rate of spindown. The Geneva model, however, shows marked differences. It lives for \qty{\sim 6}{\myr}, rather than the \qty{\sim 5}{\myr} seen in the Brott models, and loses much less angular momentum. While its initial rotation velocity is larger than that of the depicted Brott models, it spins down at a comparable rate.

\section{Inflation}
\label{app:inflation}

We have calculated a small model grid for massive MS stars with MESA, in order to investigate the effect of the convective efficiency on their envelope inflation and thus on their radius evolution \citep{sanyalMassiveMainsequenceStars2015}. Notably, envelope inflation is confirmed by 3D-radiation-hydro calculations of the atmosphere of massive stars near their Eddington-limit \citep{jiangLocalRadiationHydrodynamic2015}. We employ the standard mixing length theory for our experiment, and three different evolutionary sequences are computed per initial mass, using the canonical value for the mixing length parameter ($\alpha_\mathrm{MLT} = 1.5$), as well as one significantly smaller ($\alpha_\mathrm{MLT} = 0.5$) and one larger ($\alpha_\mathrm{MLT} = 5$) value. Seven different initial masses in the range \qty{20}{\msun} to \qty{93}{\msun} have been considered.

		\begin{figure}[!htbp]
			\centering
			\includegraphics{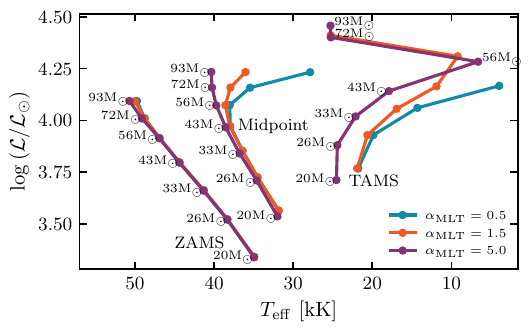}
			\caption{Spectroscopic Hertzsprung-Russell diagram showing the ZAMS line for different adopted values of the mixing length parameter $\alpha_\mathrm{MLT}$ for single star evolutionary models at $Z = 0.0092$ with $v_\mathrm{i} = \qty{150}{\kms}$. On the left are the ZAMS lines for $\alpha_\mathrm{MLT} = 0.5$ (teal), $\alpha_\mathrm{MLT} = 1.5$ (orange, the standard value), and $\alpha_\mathrm{MLT} = 5.0$ (pink). They overlap for all but the highest considered masses. The middle lines mark the midpoint of the MS, determined as the point where half of the initial hydrogen has been burnt in the core. These midpoint lines are for the same 3 $\alpha_\mathrm{MLT}$ values as for the ZAMS lines. The lines on the right mark the TAMS lines, for the same 3 $\alpha_\mathrm{MLT}$ values. The $\alpha_\mathrm{MLT} = 0.5$ line (teal) terminates at \qty{43}{\msun} and the $\alpha_\mathrm{MLT} = 1.5$ line (orange) terminates at \qty{72}{\msun}, due to the models failing to converge above this point.}
			\label{fig:mlt_zams}
		\end{figure}

As shown in Fig.\,\ref{fig:mlt_zams}, the effect of the mixing length parameter on the location of the ZAMS is small (it becomes very significant for larger initial masses; cf. Fig.\,20 in \citet{kohlerEvolutionRotatingVery2015}, and is blurred by the effect of envelope helium enrichment for the TAMS. However, the lines connecting the evolutionary mid-points, defined by models in which half of the hydrogen is burnt in the core, show a clear dependence on $\alpha_\mathrm{MLT}$ above \qty{\sim 40}{\msun}. The models with the most efficient convection remain longer in the hotter part of the sHRD than the models with less efficient convection. To the extent that the stellar wind mass loss rate depends on stellar radius of effective temperature, angular momentum loss and spindown will be affected.

				\end{appendix}
				
			\end{document}